\documentclass[aps,twocolumn,showpacs,superscriptaddress]{revtex4-1}
\usepackage{amsmath,amssymb,bm,graphicx,dsfont,color}
\usepackage[latin9]{inputenc}
\usepackage{amsmath}
\usepackage{slashed}
\usepackage{dsfont}
\usepackage{amstext}
\usepackage{amssymb}
\usepackage{amsbsy}
\usepackage{amsthm}
\usepackage{epsfig}
\usepackage{graphicx}
\usepackage{bbm}
\usepackage{hyperref}
\usepackage{color}
\usepackage{array}
\usepackage{cancel}
\usepackage{amsfonts}
\usepackage{amsbsy}
\usepackage{mathrsfs}  
\usepackage[normalem]{ulem}

\newcommand{\e}{\varepsilon}

\renewcommand{\(}{\left(}
\renewcommand{\)}{\right)}
\renewcommand{\[}{\left[}
\renewcommand{\]}{\right]}

\newcommand{\beq}{\begin{equation}}
\newcommand{\eeq}{\end{equation}}
\newcommand{\bea}{\begin{eqnarray}}
\newcommand{\eea}{\end{eqnarray}}

\newlength{\myL}

\def\be{\begin{eqnarray}}
\def\ee{\end{eqnarray}}

\begin{document}
 
\title{Universal properties of many-body delocalization transitions}
\author{Andrew C. Potter}
\affiliation{Department of Physics, University of California, Berkeley, CA 94720, USA}

\author{Romain Vasseur}
\affiliation{Department of Physics, University of California, Berkeley, CA 94720, USA}
\affiliation{Materials Science Division, Lawrence Berkeley National Laboratories, Berkeley, CA 94720}

\author{S.~A. Parameswaran}
\affiliation{Department of Physics and Astronomy, University of California, Irvine, CA 92697, USA}
\affiliation{California Institute for Quantum Emulation (CAIQuE), Elings Hall, University of California, Santa Barbara, CA 93106, USA}
\date{\today}

\begin{abstract}
We study the dynamical melting of ``hot'' one-dimensional many-body localized systems. As disorder is weakened below a critical value these non-thermal quantum glasses melt via a continuous dynamical phase transition into classical thermal liquids. By accounting for collective resonant tunneling processes, we derive and numerically solve an effective model for such quantum-to-classical transitions and compute their universal critical properties. Notably, the classical thermal liquid exhibits a broad regime of anomalously slow sub-diffusive equilibration dynamics and energy transport. The subdiffusive regime is characterized by a continuously evolving dynamical critical exponent that diverges with a universal power at the transition. Our approach elucidates the universal long-distance, low-energy scaling structure of many-body delocalization transitions in one dimension, in a way that is transparently connected to the underlying microscopic physics.
\end{abstract}

\maketitle

\section{Introduction \label{sec:intro}}
The laws of thermodynamics can break down in disordered quantum systems that are isolated from external heat sources, due to the localization of excitations that would ordinarily transport energy among distant regions to establish thermal equilibrium~\cite{FleishmanAnderson,Gornyi,BAA,PalHuse}. Such many-body localized (MBL) systems have the remarkable property that all high-energy excited states behave like
 zero-temperature quantum ground states. Their spectrum can be labeled by an extensive set of local conserved quantities~\cite{PhysRevLett.111.127201, PhysRevB.90.174202}, such that eigenstates exhibit boundary law scaling~\cite{BauerNayak} of entanglement entropy, characteristic of gapped quantum ground states. This raises the intriguing possibility that quantum coherent phenomena, typically associated with zero-temperature systems, can occur in  arbitrarily ``hot" matter. Examples of such quantum coherent phenomena include symmetry breaking below the equilibrium lower critical dimension~\cite{HuseMBLQuantumOrder}, topological edge states ~\cite{HuseMBLQuantumOrder,BahriMBLSPT, PhysRevB.89.144201,ACPMBLSPT,CenkeMBLSPT}, and quantum criticality~\cite{VoskAltmanPRL13,PekkerRSRGX,QCGPRL}. MBL systems also host novel out-of-equilibrium dynamical phase transitions, where thermodynamics breaks down sharply at a critical point~\cite{2014arXiv1407.7535B,Agarwal,Luitz,VHA}. Cold atomic, molecular, and trapped ion systems offer a promising experimental platform to explore these theoretical ideas. Indeed, issues of thermalization and excited state dynamics necessarily arise in such systems, as they are inherently well isolated from their surroundings, and typically cannot be cooled to low temperatures (compared to their characteristic energy scales). 
 
 \begin{figure}[t!]
\begin{center}
\includegraphics[width = 0.85\columnwidth]{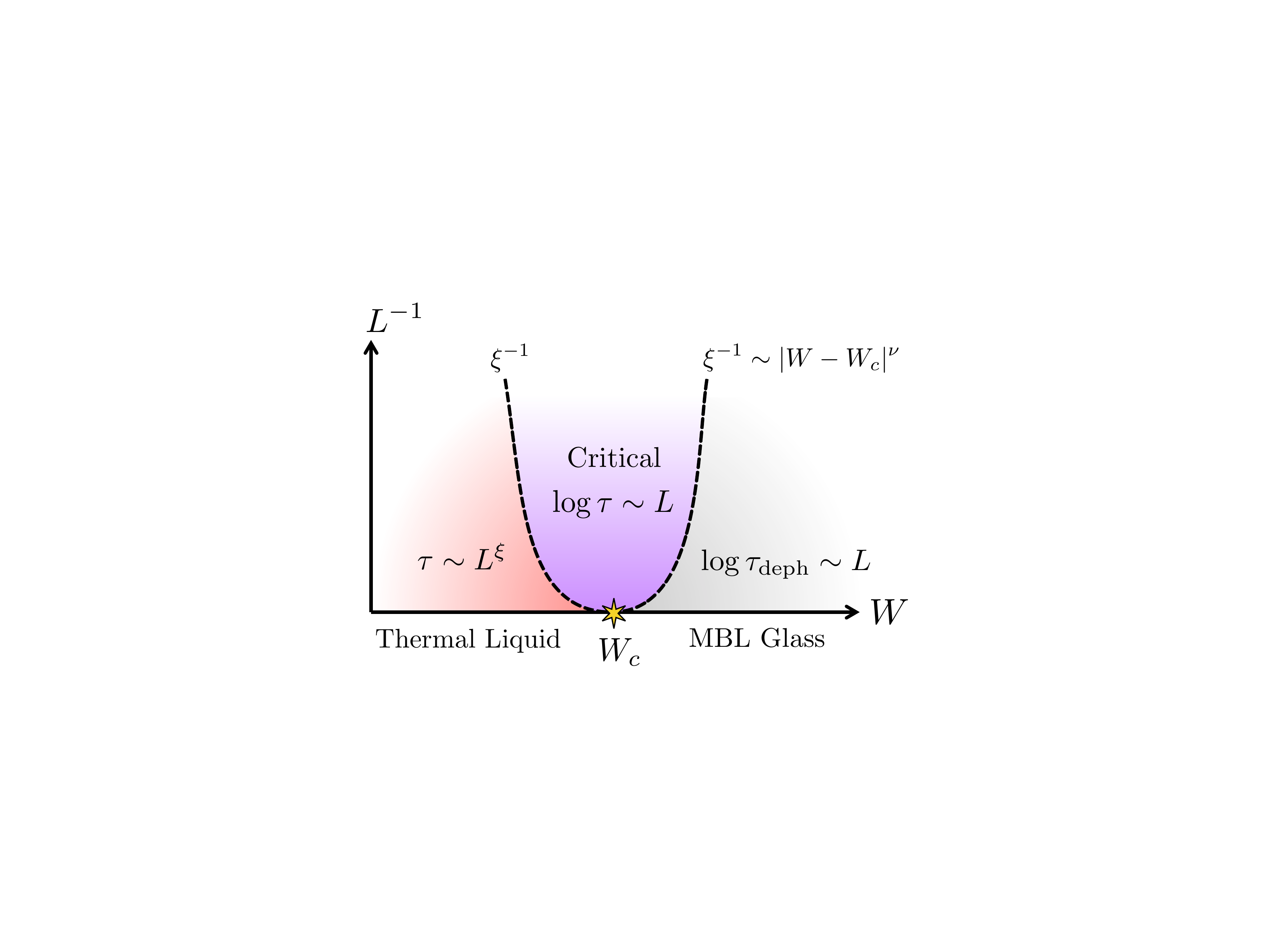}
\vspace{-.2in}
\end{center}
\caption{{\bf Phase diagram and finite-size crossovers} for one-dimensional many-body delocalization transitions. $W$ parameterizes disorder strength. For strong disorder $W>W_c$, a non-ergodic many-body localized glass is obtained, which lacks thermal transport but exhibits slow dephasing and entanglement dynamics with length-time scaling $L \sim \log  \tau_\text{\tiny deph}$ (where the subscript ``deph" emphasizes that this is the dephasing time associated with virtual quantum fluctuations rather than the energy transport time scale, which is strictly infinite in the strong disorder glass). Below a critical disorder strength, $W_c$, the MBL glass melts into a thermal, energy-conducting liquid. The melting transition is continuous (second order), characterized by a single diverging length-scale $\xi\sim |W-W_c|^{-\nu}$, where $\nu\approx 3.5\pm 0.3$. Energy transport (entanglement spreading) in the thermal liquid are characterized by a subdiffusive (sub-ballistic) power law scaling between time and length $\tau\sim L^z$, with dynamical exponent $z\sim \xi$ that diverges continuously upon approaching the transition.}
\label{fig:PhaseDiagram}
\end{figure}
 
Theoretical investigations into these questions, however, must tackle a daunting combination of out-of-equilibrium quantum dynamics, interactions, and disorder. Consequently, most existing theoretical work on MBL systems has been confined to small-scale numerics and analysis of phenomenological models. Remarkably, because of the short localization length  and short-range entanglement structure of MBL eigenstates deep within the localized phase, these approaches have met with considerable success in gleaning properties of the MBL phase itself. In contrast, these same techniques are poorly suited to access the universal properties of disordered criticality and dynamical phase transitions {\it out} of the localized phase. 

For example, whereas the ground states of disordered one-dimensional systems are always localized, the excited states are only many-body localized for sufficiently strong disorder, and melt into a self-thermalizing incoherent classical fluid at weaker disorder. These two dynamical phases are separated by an apparently continuous many-body delocalization phase transition~\cite{PalHuse}. Such excited-state delocalization transitions are neither classical thermal phase transitions nor zero-temperature quantum phase transitions, but rather, represent a novel class of dynamical quantum-to-classical criticality. Such transitions are driven by long-distance properties whose characteristic length scale diverges as the critical point is approached, rendering traditional numerical methods increasingly unreliable near the critical point.  For example, recent numerical studies~\cite{Luitz} of a many-body delocalization transition obtain critical exponents that contradict fundamental bounds~\cite{Chayes}, suggesting that small scale numerics cannot access the true long-distance scaling properties. New approaches are clearly needed to understand the nature of this new class of dynamical transitions.  

To this end, we develop an effective model for the formation of collective many-body resonances that destabilize the MBL phase at weak disorder, in order to compute the universal scaling properties of many-body delocalization transitions. While the problem of identifying generic many-body resonances is as hard as directly solving the full quantum problem, at any continuous phase transition, one expects a self-similar hierarchical scaling structure. This suggests that the critical resonant cluster may be constructed hierarchically in the spirit of the renormalization group. We implement an efficient numerical procedure to identify the formation of such delocalizing resonant clusters and study the critical properties of the resulting effective model. Our construction is motivated in part by the requirement that the MBL transition itself must be thermal~\cite{2014arXiv1405.1471G} and thus, incoherent and classical. Accordingly, the critical properties should be well described by an effective classical statistical mechanics model, {\it e.g.} which ignores quantum interference effects such as those responsible for weak-localization of non-interacting fermions in equilibrium settings. 

We obtain scaling results for the critical properties of the disorder-strength-tuned MBL transition in one-dimension, by  numerical simulation of the resonance model. We find a continuous dynamical phase transition characterized by a diverging length scale $\xi$, in which a quantum MBL glass melts into a classical thermal liquid. The predicted phase diagram and cross-over structure of this effective model are summarized in Fig.~\ref{fig:PhaseDiagram}. %

The effective resonance percolation model also provides insight into the dynamics of thermal transport and entanglement of the delocalized thermal liquid, near the melting transition (see Fig.~\ref{fig:PhaseDiagram}). In one dimension, we find that transport on the delocalized side of the transition is anomalously slow, characterized by power-law subdiffusion with a continuously evolving dynamical exponent, $z$, which diverges in a universal fashion at the MBL transition. We argue that this subdiffusion stems from tunneling through rare insulating regions, in agreement with recent numerical studies~\cite{2014arXiv1407.7535B,Agarwal} and a phenomenological renormalization group approach~\cite{VHA}, and establish a scaling relation between the correlation length exponent $\nu$, characterizing the divergence of the many-body localization length $\xi$, and the exponent governing the divergence of $z$.

This 1D MBL transition was recently studied by mean field methods~\cite{MeanFieldMBLTransition}, exact diagonalization~\cite{2014arXiv1407.7535B,Agarwal,PhysRevLett.113.107204,Luitz} and a phenomenological renormalization group procedure~\cite{VHA}. {Exact diagonalization studies on $10-20$ site chains observe an apparently continuous transition, but obtain a correlation length exponent that is incompatible with fundamental bounds~\cite{Chayes}, suggesting that the observed scaling properties are likely finite size artifacts that do not reflect the true long-distance universal properties of the MBL transition. In contrast, the critical exponents obtained in our approach comply with the Harris-Chayes bounds~\cite{Chayes}.

 Our results are broadly consistent with those of Ref.~\onlinecite{VHA}, where available. We also comment on the many-body delocalization transition of melting quantum critical glasses (QCGs)~\cite{VoskAltmanPRL13,PhysRevLett.112.217204,PekkerRSRGX,QCGPRL}, critical analogs of MBL phases that arise, for example, in random-bond spin chains. We argue that the delocalization transition for these QCG phases has identical properties to the ordinary 1D MBL transition.


\section{Model and Resonances}
Though our effective model for the MBL transition is largely insensitive to microscopic choice of a particular model, for concreteness, we will phrase our discussion in terms of a paradigmatic MBL model, the random-field XXZ chain:
\begin{align} \label{eq:InteractingAnderson}
H_\text{MBL} = \sum_i \[-J( \sigma^x_{i+1}\sigma^x_i + \sigma^y_{i+1}\sigma^y_i)- \mu_i \sigma^z_i+V\sigma^z_i\sigma^z_{i+1}\].
\end{align}
Here, $\sigma^{x,y,z}$ are Pauli-matrices for spin-1/2 degrees of freedom. This XXZ chain maps, via a standard Jordan-Wigner transformation, into a chain of spinless interacting fermions, where $J$ is the hopping amplitude, $V \ll J$ is the interaction strength, and $\mu_i$ is the chemical potential uniformly distributed on the interval $[0,W]$, where $W$ serves as a measure of disorder, and we use the spin and fermionic languages interchangeably. We will work in the Fock basis of single-particle orbitals, $\phi_i$, of the non-interacting part ($V=0$) of (\ref{eq:InteractingAnderson}), which have energy $\e_i\approx \mu_i+\mathcal{O}\(\frac{J^2}{W}\)$ and are localized near site $i$. We will then consider the interactions $V$ as a perturbation expressed in this basis, though we remark that our method is not confined to low-order perturbation theory, but rather includes contributions from all orders in $V$. We will be interested in the properties of typical highly excited states in the middle of the many-body spectrum. This Hamiltonian  has been considered as a prototypical example of a many-body localized (MBL) system (see {\it e.g.}~\cite{2014arXiv1404.0686N} for a recent review), characterized by the absence of thermal transport and by a slow, logarithmic spreading of entanglement due to dephasing~\cite{PhysRevLett.109.017202,PhysRevLett.110.260601,PhysRevB.90.174202}. When $V \neq 0$, the system is believed to exhibit a MBL-to-thermal dynamical phase transition as a function of disorder strength~\cite{2014arXiv1407.7535B,Agarwal,Luitz,VHA}, which separates a  non-ergodic MBL quantum glass phase from a ergodic high temperature classical liquid. 

To motivate the problem of delocalization, imagine creating an energy wavepacket initially localized near a single site in the MBL system. 
How does such a packet evolve with time?
Deep in the localized phase, it is extremely unlikely that it 
can tunnel from a given site (bond) to another, due to the wide variance of site (bond) energies. More precisely, the amplitude to tunnel through a distance $x$ of the localized phase is
\begin{align}
\Gamma(x) \approx Ve^{-x/x_0},
\label{eqJtypgen}
\end{align}
%
where $x_0 \approx \[\frac{1}{2}\ln\(1+\(\frac{W}{J}\)^2\)\]^{-1}$, is the single particle localization length (measured in units of the lattice spacing), where we have chosen an expression that interpolates smoothly between the strong disorder limit $x_0\approx \ln^{-1}\frac{W}{J}$, and weak disorder limit $x_0\approx  \(\frac{J}{W}\)^2$. If the tunneling strength, $\Gamma(x_{ij})$, between spin pairs $i$ and $j$ separated by distance $x_{ij}$ is much smaller than their energy difference $\delta E_{ij}$, then the true many-body eigenstate is very close to a product state of independent configurations of each bond. On the other hand, if $\Gamma(x_{ij})\gtrsim \delta E_{ij}$, then the many-body eigenstate contains an entangled superposition of the degrees of freedom on the two sites, and we say that the two sites are {\it resonantly linked}. We estimate the probability that a given site is resonantly linked to at least one other is $(1-e^{-\lambda_R})$, with $\lambda_R = \nu(0)\int_0^\infty dx \Gamma(x)\approx \frac{V}{J}x_0$, where $\nu(0)\approx 1/J$ is the single particle density of states.

For very strong disorder ($x_0\ll 1$), the density of resonantly linked pairs is very small, $\rho_\text{res} \approx \lambda_R\ll 1$, so that the resonant links are well isolated from each other, and they do not disrupt the properties of the surrounding localized phase. As disorder is weakened, these resonant links occur more frequently, and eventually disrupt localization below a critical disorder strength $W_c$.  Since, in a thermal phase, all spins are highly inter-entangled, delocalization and the recovery of thermalization at weak disorder must occur via a highly collective resonance involving every single spin. Identifying all possible generic many-body resonances is a hard problem, and in fact as difficult as exactly solving the full quantum many-body problem. 
However, near the delocalization critical point, it is natural to expect that the critical resonant cluster exhibits scaling structure that is geometrically self-similar (in a statistical sense) at different energy scales. Anticipating this hierarchical structure motivates a simpler way of identifying collective resonances by first identifying small resonant clusters (e.g. resonant pairs as above), and then examining whether groups of these small resonant clusters can collectively resonate on some lower energy scale (longer timescale), and so on. Using this perspective, we will construct an effective quantum percolation model for delocalization by hierarchical collective resonances. 

\section{Effective model for the delocalization transition}
A cluster of $m$ resonantly linked sites is characterized by bands of $\mathcal{N} \approx 2^{m}$ many-body energy levels, and thus has an exponentially small level spacing in $m$. Roughly speaking, collective rearrangements among spins of two such clusters containing $m_{1,2}$ sites respectively and separated by distance $L_{12}$, can occur resonantly if the effective tunneling $\Gamma_{12}$ is larger than the level spacing of the combined cluster, scaling as $\sim 2^{-(m_1+m_2)}$. We therefore see that this exponential suppression of the level spacing will favor large, many-spin resonant clusters. This motivates an effective model for the delocalization transition, consisting of the following iterative procedure:
\begin{enumerate}
\item Examine a disordered chain to identify resonantly linked pairs of spins, and group them into 2-spin clusters
\item Identify the strongest resonantly coupled pair of clusters (i.e. the one with largest interaction $\Gamma_{ij}$) and merge them into a new larger resonant cluster
\item Account for the reduction of level spacing on newly formed resonant cluster
\item Compute the effective interaction of the new cluster with other clusters (as described in detail below)
\item Examine whether the reduced level spacing enables any of the newly formed clusters to resonantly interact, if so merge these clusters  
\end{enumerate} 
Steps 2-5 can then be iterated (numerically) until all possible clusters are formed. 

Before giving a detailed account of this procedure, we briefly sketch its potential outcomes and underlying philosophy. For strong disorder, the resonance merging ceases after a small number of steps, producing only dilute well isolated small resonant clusters embedded inside strongly localized quantum glass. Upon weakening disorder, the size of the resonant clusters increases, diverging continuously at a critical value of disorder $W=W_c$, where a single cluster just barely percolates across the system (in the limit of infinite system size). For weaker disorder ($W<W_c$), the resonance merging procedure continues until all degrees of freedom are subsumed into a single percolating cluster (in the infinite system limit), resulting in a delocalized thermal phase. Nevertheless, we will see that the near critical transport properties of the delocalized phase exhibit anomalously slow dynamics, with an unusual scaling structure, due to the proximity to the quantum glass phase.

In this procedure, resonances arise from small mismatch in energy scales, corresponding to a slow beating frequency or long time in dynamical processes. Over the coarse of iterations, the dynamical time scales associated with the merging of resonances steadily lengthen. In this way, the effective model closely resembles a renormalization group procedure -- first accounting for fast degrees of freedom and progressively proceeding to slower and slower modes. We also remark that, in principle, the problem of identifying generic many-spin resonances is factorially complicated in system size, and hence is no simpler than directly solving the full quantum problem. From our analytic treatment of the strong disorder phase, we expect that the most likely type of resonances are those between two bonds (orbitals), motivating our use of step 1) above to ``seed" the iterative resonance merging procedure. Moreover, near the transition, whether or not a resonant cluster percolates across the system is driven by highly collective resonances among large clusters of many spins, and hence the low-energy, long time scale dynamics of the near-critical cluster are expected to be insensitive to the details of early steps in the procedure (i.e. are universal).

\subsection{Criterion for merging resonant clusters}
To test if two clusters $i$ and $j$ can resonate (step 2), we compare their coupling $\Gamma_{i,j}$ to the energy mismatch, $\delta E_{i,j}$ of their energy levels. For two large clusters $i,j$, such that the bandwidth of each cluster exceeds the level spacing of the other ($\min\{\Lambda_i,\Lambda_j\}>\max\{\delta_i,\delta_j\}$), we define an energy mismatch $\delta E_{i,j}$ between two clusters $i,j$ as: $\frac{\delta_i\delta_j}{\min\{\Lambda_i,\Lambda_j\}}$. In the other case, where one cluster's bandwidth is smaller than the other's level spacing, we define $\delta E_{i,j} = \max\{\delta_i,\delta_j\}-\min\{\Lambda_i,\Lambda_j\}$ (see Appendix~\ref{SecResMergingApp}). We take the bandwidth of the newly merged cluster to be: $\Lambda_{i\cup j}=\Lambda_i+\Lambda_j+\Gamma_{i,j}$, and the level spacing to be $\delta_{i\cup j} = \frac{\Lambda_{i\cup j}}{2^{m_i+m_j}-1}$. We emphasize that the main ingredient at this step is the exponential reduction of the level spacing when two clusters are merged: we checked numerically that the universal properties of the transition are insensitive to the other details of the procedure.

\subsection{Renormalization of inter-cluster couplings}
After merging two clusters, one should in principle compute a new set of inter-cluster couplings (step 4 above). Consider merging two clusters $i$ and $j$, with number of spins $m_{i},m_{j}$ and bandwidths $\Lambda_{i},\Lambda_{j}$ respectively, and interacting with coupling $\Gamma_{i,j}$. In order for a third cluster, $k$, to drive a collective rearrangement of all $2^{m_i+m_j}$ levels of the merged cluster, $i\cup j$, and hence access the full many-body level spacing of $i\cup j$, $k$ must interact separately with both $i$ and $j$. The effective interaction strength for such a process depends on the energy mismatch between $k$ and $i,j$ respectively, and can occur by several different possible processes. For instance, consider the process $k\rightarrow i \rightarrow j$, where $k$ first interacts with $i$ and then $i$ with $j$. If the energy mismatch, $\delta E_{k,i/j}$ between $k$ and $i$ or $j$ is large compared to the couplings $\Gamma_{k,i/j}$ {between $k$ and $i$ or $j$ }, then the interaction between $k$ and $i\cup j$ can be computed perturbatively:
\begin{align}
\Gamma^{{(k\rightarrow i\rightarrow j)}}_{k,i\cup j} \approx \frac{\Gamma_{k,i}\Gamma_{i,j}}{\delta E_{k,i}}.
\end{align}
In the alternative case where all three clusters $i$, $j$, and $k$ can resonate , e.g. $\Gamma_{k,i}>\delta E_{k,i}$, 
we may estimate the effective coupling between $k$ and $i\cup j$ by
 classically adding the time taken for an excitation to transfer from $k\rightarrow i\rightarrow j$:
\begin{align}
\Gamma^{{(k\rightarrow i\rightarrow j)}}_{k,i\cup j} \approx \(\Gamma_{k,i}^{-1}+\Gamma_{i,j}^{-1}\)^{-1}.
\end{align}
There are a few possible routes for $k$ to excite both $i$ and $j$ -- for example $k$ can first excite $i$, then $i$ excites $j$, or $k$ can excite both $i$ and $j$, etc. We compute $\Gamma_{k, i\cup j}$ for each process using the `perturbative' or `classical' rules above as appropriate, and choose their maximum as the new effective coupling between $k$ and $i\cup j$.

Computing the renormalized inter-cluster couplings in step 4 is conceptually important to avoid a potential instability of the strong disorder phase (see Appendix~\ref{AppModel}), however, in practice we find that this instability is absent for all numerically accessible system sizes and find identical scaling independent of whether or not the inter-cluster couplings are renormalized. We remark that the way the couplings are renormalized is reminiscent of the approach of Ref.~\cite{VHA} where metallic and isolating blocks are merged by using a renormalization group scheme. However, we point out that we only merge ``metallic'' (resonant) clusters in our model, thus avoiding the thorny issue of determining the proper renormalized couplings after merging ``mixed'' insulating and thermalizing blocks.

\section{Scaling Structure of the MBL Transition}

\begin{figure}[t!]
\begin{center}
\includegraphics[width = 1.0\columnwidth]{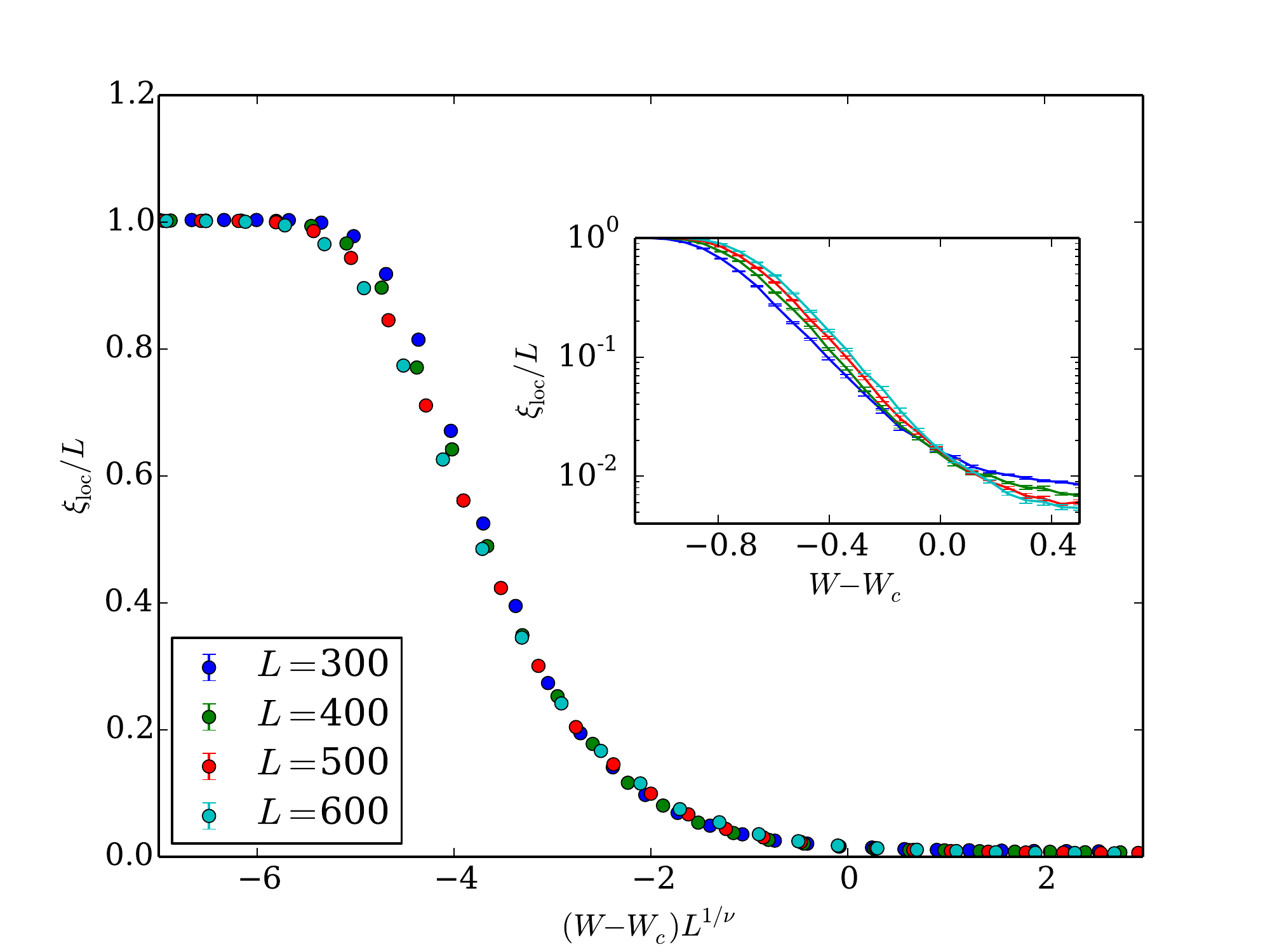}
\end{center}
\caption{{\bf Finite-size scaling for the disorder averaged localization length $\xi_{\text{loc}}$} (defined as the size of the longest resonant cluster for a given disorder realization). $\frac{\xi_{\text{loc}}}{L}$ collapse to a universal form (main panel) when plotted against the scaling variable $(W-W_c)L^{1/\nu}$, with correlation length exponent $\nu \approx 3.5 \pm 0.3$. Insets show the unscaled $\frac{\xi_{\text{loc}}}{L}$ curves. Results are averaged over $\sim 10^4$ disorder realizations. 
}
\label{fig:XiScaling}
\end{figure}

To identify the scaling structure of the transition predicted by the effective resonance percolation model described above, we examine the dependence of the size, $\xi_\text{loc}$, (number of spins) of the largest resonant cluster as a function of disorder strength, $W$, for various system sizes, $L$. Disorder averaged results for $V=0.3$ are shown in Fig.~\ref{fig:XiScaling}. For $W\ll W_c$ one obtains a percolating cluster with $\xi_\text{loc} \approx L$ with probability close to one. In the finite size systems simulated, $\xi_\text{loc}/L$ crosses over from one ($W\ll W_c$) to zero ($W\gg W_c$) as disorder is increased. 

The curves for different system sizes cross at the same value of disorder strength, which we identify as the critical disorder strength $W_c$. Moreover, the data for different system sizes collapse to a universal scaling form $\xi_\text{loc} = L~ \Xi\((W-W_c)L^{1/\nu}\)$, with correlation length exponent $\nu = 3.5 \pm 0.3$, as shown by plotting $\frac{\xi_\text{loc}}{L}$ against the scaling variable $(W-W_c)L^{1/\nu}$ (Fig.~\ref{fig:XiScaling}). Here $\Xi(x)$ is a universal scaling function interpolating between one $(x\ll 0)$ and zero $(x\gg 0)$.

This scaling collapse indicates the existence of a length scale $\xi$ that diverges when the transition is approached from either side as $\xi \sim |W-W_c|^{-\nu}$, signaling a sharp continuous delocalization phase transition. The diverging length scale $\xi$ can be interpreted as the localization length for $W>W_c$, and as we will see below, characterizes the length of insulating gaps in the transport path for $W<W_c$. 

We remark that the value of the scaling function at criticality: $\Xi(0)\approx 10^{-2}$, is anomalously low compared to ordinary percolation (for which $\Xi(0)$ would be $\approx 0.5$). This strong asymmetry indicates that the transition is driven by rare resonant clusters that are quite sparse and widely separated (e.g. clusters of size $\ell$ are separated by typical distance $\approx \Xi(0)^{-1}\times \ell\gg \ell$ at criticality).

We also performed simulations of a simpler model that ignores the renormalization of the couplings when merging clusters. As claimed above, despite the potential stability issues of the strong disorder phase within that simplified model, we find essentially identical universal percolation curves using these simplified rules, compatible with the same universal exponent $\nu \approx 3.5$ within error bars (see Appendix~\ref{AppModel}). 

\section{Near-critical dynamics in the delocalized phase}
We now turn to the task of computing the near-critical scaling of transport and entanglement dynamics, in the delocalized phase ($W\lesssim W_c$). Our goal will be to understand the scaling structure underpinning the critical slowing down of energy transport as $W$ approaches $W_c$ from below. Since the delocalized phase is thermal and at high temperature~\cite{2014arXiv1405.1471G}, transport occurs via the thermally incoherent transfer of energy among bonds in the resonant cluster. Hence, we model the spread of excess energy (initially localized near position $x$ as a function of time) as a classical random walk across the resonant cluster, with a timescale $\tau_{AB}=1/J(L_{AB})$ to transfer excitations between resonantly coupled bonds $A$ and $B$ separated by distance $L_{AB}$. 

The iterative merging procedure yields detailed information about the connectivity structure and the coupling strengths for each participating link in the resonant cluster. Here, it is important to note that, while all degrees of freedom are part of the percolating resonant cluster for $W<W_c$, not all spins contribute equally to transport and dynamics. Rather, dynamical properties are dominated by a subset of efficiently connected spins that form the ``backbone" of the transport path through the resonant cluster. Then, using standard Green's function methods, we compute the time evolution of an initially well-localized energy wave packet spreading across the resonances network via a random walk. Our main result is that the delocalized phase exhibits a broad regime of anomalously slow sub-diffusive equilibration dynamics and energy transport. We observe this subdiffusive behavior in our model using both renormalized and non-renormalized couplings, but find that finite-size effects are much weaker for the simplified model that ignores renormalization, thus allowing us to extract much cleaner data. We will therefore focus on this simplified model in the following -- recall that it reproduces equally well the critical percolation curves described in the previous section~\footnote{Properly identifying the transport path in this simplified model requires some care, see Appendix~\ref{app:details} for details.}.

\begin{figure}[t!]
\begin{center}
\includegraphics[width = 1.0\columnwidth]{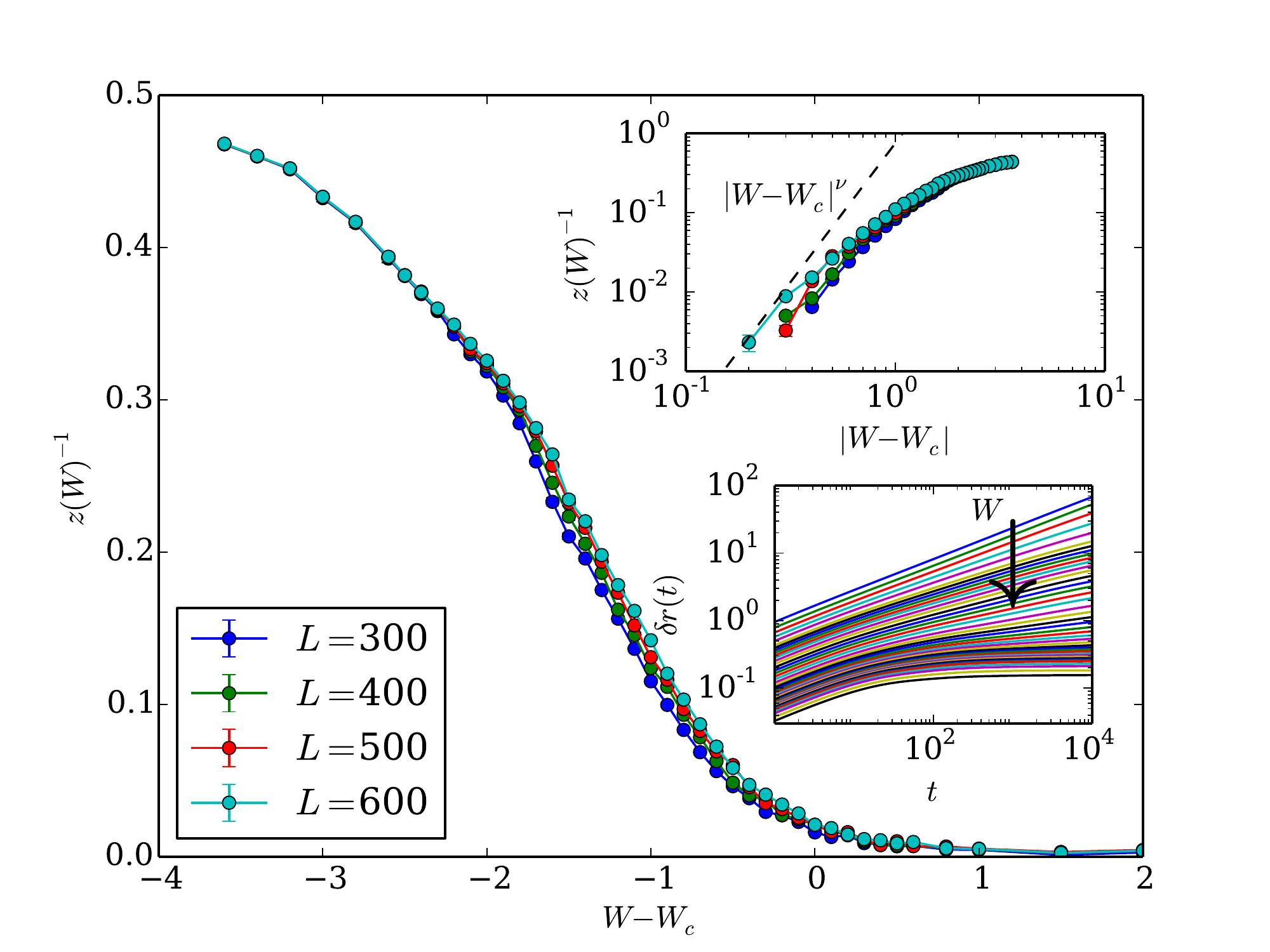}
\end{center}
\caption{{\bf Energy transport near the transition.} In the delocalized phase ($W<W_c$) excess energy initially localized near the origin spreads to distance $\delta r\sim t^{1/z}$ in time $t$ (bottom inset). The dynamical critical exponent $z$ diverges continuously from the delocalized side of the transition, and vanishes inside the quantum critical glass phase (main panel). The numerical results are consistent with $z$ diverging as $|W-W_c|^{-\zeta}$ with $\zeta = \nu$ (top inset).}
\label{fig:Transport}
\end{figure}

Averaging over initial position $x$, and over disorder configurations, we find that the mean-square displacement of an excitation grows as a power~law in time (Fig.~\ref{fig:Transport}),
\begin{align} \overline{|\delta x(t)|} \approx t^{1/z(W)}.
\label{eq:Subdiffusion}
\end{align}
For ordinary classical diffusion, 
 $z = 2$. In contrast, we find that $z(W)$ diverges in the glassy phase, indicating an absence of energy transport and a breakdown of thermal equilibration. On the delocalized side of the critical point, $1/z(W)$ increases continuously from zero as a function of detuning from the critical disorder strength:
\begin{align}
z(W) \approx \frac{z_0}{\(W_c-W\)^{\zeta}}, \hspace{.2in}(W\lesssim W_c).
\label{eq:z}
\end{align}
Though seemingly a distinct universal exponent, $\zeta$ is related by a general scaling relation to the log-dynamical exponent $\psi$ of the localized phase, and the correlation length exponent $\nu$. To see this, note that at the critical point, transport along the critical resonant chain occurs by tunneling through arbitrarily long localized regions, hence energy scales with length as $\sim e^{-L}$ at the critical point, just as in the localized phase. This must cross over smoothly to the power law scaling of (\ref{eq:Subdiffusion}) on length scales $L\approx \xi$, implying the scaling relation:
\begin{align}
\zeta = \nu.
\end{align}
Using this scaling relation and the definition of the correlation length exponent we may rewrite (\ref{eq:z}) as $z\sim \xi$  (ignoring logarithmic corrections and non-universal prefactors). Accessing the scaling regime~(\ref{eq:z}) to measure $\zeta$ numerically is difficult because of the form of the finite-size percolation curves (Fig.~\ref{fig:XiScaling}): close to the critical point with $W<W_c$, $\xi_{\rm loc} > L$ even for moderately large systems. In particular, the small non-zero value of $z^{-1}$ at $W_c$ is a finite size artifact, and upon subtracting the value of $z(W_c)^{-1}$, we find critical behavior compatible with the expected scaling relation $\zeta=\nu$ (Fig.~\ref{fig:Transport} upper inset).   

\subsection{Structure of transport path}
Subdiffusive transport arises from the broad distribution of effective tunneling links, $\Gamma_{ij}$, in the resonant backbone, corresponding to a power-law distribution of timescales~\cite{Hulin,Agarwal}, $\tau_{ij}\sim 1/\Gamma_{ij}$:
\begin{align}
p(\tau) \sim \frac{1}{\tau^{\alpha}}.
\label{eqWaitingTimes}
\end{align}
with $1<\alpha \leq 2$ such that the mean waiting time $\int_{0}^{\infty} \tau p(\tau) d\tau $ is divergent. To see the relation between this power-law distribution and  subdiffusive transport properties, note that energy transport occurs as a random walk on a one dimensional chain whose ``sites" are two-spin bonds in the resonant cluster, and whose links are weighted by a waiting time $\tau$ corresponding to the inverse effective coupling between ``sites" (In principle the bonds also have variable lengths, $\ell\sim \log \tau$. However, unlike the waiting-times the bond-lengths have a finite mean, $\bar{\ell}$, and in the following we approximate the length of each step as $\bar{\ell}$). In $N$ steps, a random walker traverses 
$\sim\mathcal{O}(\sqrt{N})$ different bonds, revisiting each approximately $\sqrt{N}$ times, and moving total distance $L \approx \sqrt{N}$ (in units of the average step size).  For broad distribution of waiting times, the time taken for $N$ steps is dominated by the longest waiting time, $T$, encountered. The probability of encountering a bond with waiting time of at least $T$ is $P(\tau\geq T) = \int_T^\infty p(\tau)\approx T^{1-\alpha}$. 
We can reasonably expect to find a link with waiting time $\tau\gtrsim T$ among $L\approx \sqrt{N}$ different bonds if $LP(\tau\geq T)\approx 1$, 
and therefore the slowest link in  region of size $L$ has waiting time $T(L)\approx L^{\frac{1}{\alpha-1}}$. Since this weak link is revisited order $\sqrt{N}\approx L$ times, the total time to move distance $L$ scales as $t_\text{energy}(L)\approx L^{\frac{\alpha}{\alpha-1}}$. Comparing to 
(\ref{eq:Subdiffusion}), we identify the dynamical exponent for energy transport as:
\begin{align} 
z\equiv z_{\rm energy}=\frac{\alpha}{\alpha-1} ,
\end{align}
implying that $\alpha$ approaches one near the transition as: $\alpha(W)-1 \sim |W-W_c|^\zeta$.

Note also that the broad distribution of waiting times~(\ref{eqWaitingTimes}) corresponds to a probability of encountering a link of length $\ell = \log \tau$ in the transport path along the resonant backbone, near the transition, 
 given by 
\begin{align} P(\ell) \sim e^{(1-\alpha) \ell} \sim e^{ -l/z} =e^{- \(\frac{\ell}{\xi}\)}.
\end{align}
Hence the typical spacing between ``vertebrae" in the resonant backbone is $\approx \xi$, corresponding to energy scale $e^{-\xi}$. However, in a region of length $L\gg \xi$, it is extremely unlikely to avoid encountering an atypically long gap of length $\ell_\star$, defined by $\frac{L}{\xi}P(\ell_\star)\sim 1$, i.e. $\ell_\star \sim \xi \(\log \frac{L}{\xi}\)$. Such long rare links involve waiting time $\tau \sim  e^{l_{\star}} \sim L^{\xi}$, which dominates the time required to traverse a segment of size $L$.

\subsection{Entanglement dynamics vs. thermal transport}
We note that the dynamics of energy transport scale differently than those of entanglement since energy is a conserved quantity that can only be transferred among subregions whereas entanglement can be freely generated~\cite{KimPRL13,VHA}. Hence, whereas energy transport can be viewed as a random walk of conserved particle-like excitations, entropy spreads deterministically in all directions simultaneously~\cite{KimPRL13}. In order to entangle regions separated by distance $L$ , entanglement must spread across order $L$ links, only visiting each once (in the same spirit as the second law of thermodynamics, once a bond is entangled with many others, it is extremely unlikely to later disentangle itself). Therefore the timescale for entanglement spread is dominated by the longest typical waiting time encountered, 
 $t_\text{ent}\approx T(L)\approx  L^{\frac{1}{\alpha-1}}$. Comparing, we find a different effective dynamical exponent for the spread of entanglement:
\begin{align}
z_\text{ent} = \frac{1}{\alpha-1} = z_\text{energy}-1,
\end{align}
with $z_\text{ent} \sim z \sim \xi$ near the transition.

\subsection{Scaling of optical conductivity}
So far we discussed transport from the perspective of the propagation of an initially well-localized energy wave packet. This is natural in ultracold atomic systems (see {\it e.g.}~\cite{PhysRevLett.110.205301}). The power-law distribution of time scales also implies an anomalous power-law frequency dependence for AC conductivities. For systems with a conserved spin-component or particle number, one can compute the scaling of the optical number conductivity via the Einstein relation $\sigma(\omega)\approx D(\omega)\chi$, where $D(\omega)$ is the (frequency-scale dependent) diffusion constant, and $\chi$ is the static compressibility (which will be constant throughout the phase diagram). The diffusion constant scales like $D\sim \omega L^2\sim \omega^{1-2/z}$ where $z=z_\text{energy}$ is the dynamical scaling of energy, hence the optical conductivity will also scale as:
\begin{align}
\sigma(\omega)\sim \omega^{1-2/z}.
\end{align}
This expression interpolates between constant conductivity in the diffusive limit ($z=2$) for weak disorder, to $\sigma(W_c)\sim \omega$ at the MBL transition $z=\infty$. This scaling argument agrees with the recent rare-regions analysis and numerical results of Ref.~\cite{Agarwal}.

In two dimensions, we expect a non-zero diffusion constant and finite DC conductivity in the entire region $W<W_c$. By the above argument, our effective resonance model again predicts $z=\infty$ (tunneling-like) dynamics at the MBL transition in higher dimensions, suggesting $\sigma(\omega)\sim \omega$ in the critical regime.


\section{Summary of results}
\subsection{Phase diagram}
Let us summarize our results in the thermal phase ($W < W_c$). Near the critical point, the system exhibits anomalously slow subdiffusive dynamics $\tau \sim L^{\xi}$. Inside the critical regime (i.e. on length scales $L\ll |W-W_c|^{-\nu}$), the critical resonant cluster mediates coherent transport of energy with characteristic scaling of length and time $\log \tau \sim L$. This scaling relation is reminiscent of the one governing the dynamics of dephasing and entanglement growth in the localized MBL phase, with the important distinction that dynamics in the strong disorder MBL phase describe only {\it virtual} (``off-shell") transitions rather than {\it real} (``on-shell", or resonant) processes required for thermal transport. Hence, whereas the MBL glass is non-ergodic, the energy eigenstates at the glass melting critical point are ergodic and thermal, with volume law scaling of entanglement in every eigenstate (consistent with general entanglement monotonicity requirements~\cite{2014arXiv1405.1471G}). However, while the pure eigenstates at the delocalization transition exhibit thermal behavior, starting from a superposition of energy eigenstates (the only initial conditions that can be prepared experimentally in finite time), the system will take super-polynomially long time (in system size) to equilibrate and thermalize. 
The resulting phase diagram and crossover scales are shown in Fig.~\ref{fig:PhaseDiagram}.

\subsection{Quantum critical glasses and extended interactions}
We now briefly comment on the generalization of our results to  critical analogs of MBL systems, dubbed quantum critical glasses (QCGs) in Ref.~\onlinecite{QCGPRL}. Examples of QCGs include phase transitions between MBL phases with different (discrete) symmetry or topological order~\cite{VoskAltmanPRL13,PhysRevLett.112.217204,PekkerRSRGX,QCGPRL}, as well as stable excited state critical phases~\cite{QCGPRL}. At strong disorder, QCGs are characterized by a stretched exponential tunneling in which energy, $E$, scales with distance, $x$, as: $E\sim e^{-(x/x_0)^\psi}$, with universal exponent $\psi<1$ whose value depends on the universality class of the QCG in question. As for the MBL case, upon weakening disorder, one expects a continuous delocalization transition from QCG to thermal liquid in which resonant clusters percolate. 

Whereas the renormalization of inter-cluster couplings plays little role in the scaling properties of the MBL transition, we expect it may play a more crucial role in the QCG delocalization transition. Loosely, two clusters with number of spins $n_{1,2}$ respectively, can inter-resonate if their coupling $\Gamma_{12}$ exceeds the many-body level spacing on the clusters, $\approx 2^{-(n_1+n_2)}$. As the clusters are formed, the bare coupling between them, $\Gamma_0(x_{12})\approx e^{-(x_{12}/x_0)^\psi}$ (where $x_{12}$ is the inter-cluster distance), is perturbatively reduced by a factor of $\approx \(\frac{e^{-(x_\text{typ}/x_0)^\psi}}{\Delta_\text{typ}}\)^{n_1+n_2}$, where $x_\text{typ}$ is the typical spacing of spins in the clusters, and $\Delta_\text{typ}$ is their typical energy mismatch. At the delocalization transition, one expects the clusters to just barely resonate, i.e. $n_1+n_2\approx x_{12}$, giving effective inter-cluster coupling $\Gamma_{12}\approx e^{-x_{12}/\tilde{\ell}}\Gamma_0(x_{12})$, where $\tilde{\ell}^{-1}\approx (x_\text{typ}/x_0)^\psi-\log \Delta_\text{typ}$. As the exponential prefactor dominates over the stretched exponential bare coupling $\Gamma_0$, this indicates the renormalization of couplings gives effective exponential-in-distance interactions between large resonant clusters in QCGs near their delocalization transition, in contrast to their longer range stretched-exponential-in-distance interactions between individual spins. This suggests that the QCG delocalization transition is actually in the same universality class as the MBL transition. The above scenario contrasts an alternative picture of the QCG-delocalization transition as being driven chunks of perfect metal interacting through QCG regions via stretched-exponential-in-distance couplings, which would predict an absence of subdiffusion in the delocalized phase \cite{VHA}. The above considerations instead suggest that the ``metallic" clusters in this picture are better thought of as ``critical" clusters, and interact with highly renormalized exponential-in-distance interactions. 

We note that similar arguments can be used to suggest the stability of QCGs with stretched exponential interactions, MBL phases with (short-ranged) power-law interactions, $V(x)\sim \frac{1}{x^p}$, and MBL in higher dimensions for strong disorder. Loosely, though the two-spin interaction, $V(L)$ between spins of scale $L$, is much larger than the many-body level spacing $\delta(L)\sim 2^{-L^d}$ (where $d$ is the number of spatial dimensions) for large $L$, in order to access the small many-body level spacings, $\sim L^d$ powers of the two-spin interactions must be used. For strong disorder, each of two body transition is energy-mismatched such that the many-body matrix element, $\Gamma(L)$ corresponding to transitions with level spacing $\delta(L)$ also scales as $\Gamma(L)\sim \(\frac{V}{W}\)^{L^d}$ allowing for an MBL phase for $W\gg V$. Similar arguments to those above suggest that the MBL transition with extended interactions will be of the same universality class as that with strictly short-ranged interactions -- due to the renormalization of the interactions between resonant clusters. We leave a more detailed study of these issues for future work.

\subsection{Validity of quantum percolation model}
Before concluding, we pause to comment on the assumptions and approximations that have gone into constructing the effective quantum percolation model of the delocalization transition. {Our key assumption is that the critical resonant cluster has a self-similar structure in energy and space, which allows us to hierarchically construct the percolating resonance that drives delocalization. This hierarchical construction clearly does not describe the most general collective resonances, and in particular ignores generic unstructured resonant clusters that do not decompose into a hierarchical tree of few-body resonances. However, the assumption of self-similarity is naturally expected for a continuous phase transition characterized by a diverging length scale, based on knowledge of more conventional disordered criticality. We therefore expect our approach to accurately capture the universal aspects of 
such a resonant structure close to the delocalization transition.

{Our approach also seeks to construct a delocalizing resonance by merging metallic clusters. Such an approach is known to be problematic for non-interacting Anderson insulators in which the conductivity of blocks does not add in a simple fashion due to coherent interference effects~\cite{LocalizationResistanceAdditionSubtleties}. However, in the present context, since the system must be thermal at the MBL transition, we expect strong dephasing to wash out any effects of quantum interference (e.g. the time-scale for dephasing is more rapid than that of transport, and does not diverge even in the MBL phase).}

\section{Discussion\label{sec:Discussion} }
To summarize, we have developed an effective model for the many-body delocalization transition of a one-dimensional MBL phase. This model can be efficiently simulated for very large system sizes (with computational cost scaling polynomially in system size), and we argue that it gives an accurate description of its long-distance, low-energy, universal critical properties. 

This transition represents a novel form of non-equilibrium excited-state criticality, distinct from classical, thermal phase transitions and zero-temperature quantum-criticality. We find that this excited state transition is continuous (second order), characterized by a single diverging length scale. We describe the set of universal critical exponents that characterize this melting transition, which are connected by scaling relations.  In particular, we find an unusual scaling structure of the critical slowing down of dynamics and thermal transport characterized by a continuously evolving dynamical exponent, $z$, which diverges in a universal fashion upon approaching the quantum critical point. This unusual, universal divergence of a critical exponent, $z$, stems from a sparse structure to the transport path through the critical percolating cluster. It is natural to expect that this sparsity corresponds to a (multi)-fractal structure of the critical cluster. While our approach makes many simplifying approximations, and is clearly not exact, we believe the general scaling structure it predicts (Fig.~\ref{fig:PhaseDiagram}) to be correct, and moreover expect to obtain reasonable (though possibly approximate) value for the correlation length exponent, $\nu$. In particular, it is encouraging that the scaling data of Fig.~\ref{fig:PhaseDiagram} is closely compatible with the renormalization group approach of Ref.~\onlinecite{VHA}~\cite{HusePC}, despite the marked differences between these approaches. 
Testing these results using standard numerical techniques may be challenging. In particular, a direct frontal assault via exact diagonalization is unlikely to give accurate scaling results in systems of only a few tens of sites.

We anticipate that these methods can be extended to treat a variety of other out-of-equilibrium dynamical phase transitions in disordered quantum systems. For instance, our analysis in terms of quantum percolation of resonant clusters can be straightforwardly adapted to treat higher dimensional delocalization transitions, which are so far inaccessible by other methods. One also expects a self-similar fractal structure of the critical resonant cluster, which may have implications for certain scaling properties and many body level statistics at the delocalization transition. A key expected difference in higher dimension is that the near-critical transport will not be dominated by rare insulating gaps, which can be simply circumvented in higher dimensions. This suggests that the delocalized thermal liquid will be diffusive with dynamical exponent $z=2$. Then repeating the arguments leading to the scaling relation $\zeta=\nu$ above, one would expect the diffusion constant to vanish as $D\sim e^{-\xi/x_0}$ upon approaching the higher-dimensional MBL transition.

Another unanswered question regarding one-dimensional systems concerns the fate of subdiffusion deep in the delocalized phase at weak disorder, far from the MBL transition. One possible scenario is that subdiffusion persists to arbitrarily weak disorder, such that diffusion is only asymptotically recovered in the clean limit. This would imply that any amount of disorder in one-dimensional systems leads to subdiffusive dynamics. The alternative would a second finite-disorder-strength excited state phase transition separating diffusive and subdiffusive regimes. We leave a detailed investigation of such issues for future work.\\

 \vspace{12pt}\noindent{\it Acknowledgements.- } We are grateful to D. Huse and E. Altman for many insightful comments and a detailed comparison of results of Ref.~\cite{VHA}. We also acknowledge helpful conversations with J.E. Moore, A. Vishwanath, R. Vosk, C.R. Laumann and S.L. Sondhi. This work was supported by the Gordon and Betty Moore Foundation's EPiQS Initiative through Grant GBMF4307 (ACP), the Quantum Materials Program at LBNL (RV), and UC Irvine startup funds (SAP).
\bibliography{MBL}

\begin{appendix}
\section{Details of Effective Model of the Delocalization Transition}\label{app:details}
\setcounter{figure}{0}
\makeatletter 
\renewcommand{\thefigure}{C\@arabic\c@figure}
\makeatother

\label{AppModel}

\subsection{Resonance Merging Procedure}

\begin{figure}[b!]
\begin{center}
\includegraphics[width = 0.7\columnwidth]{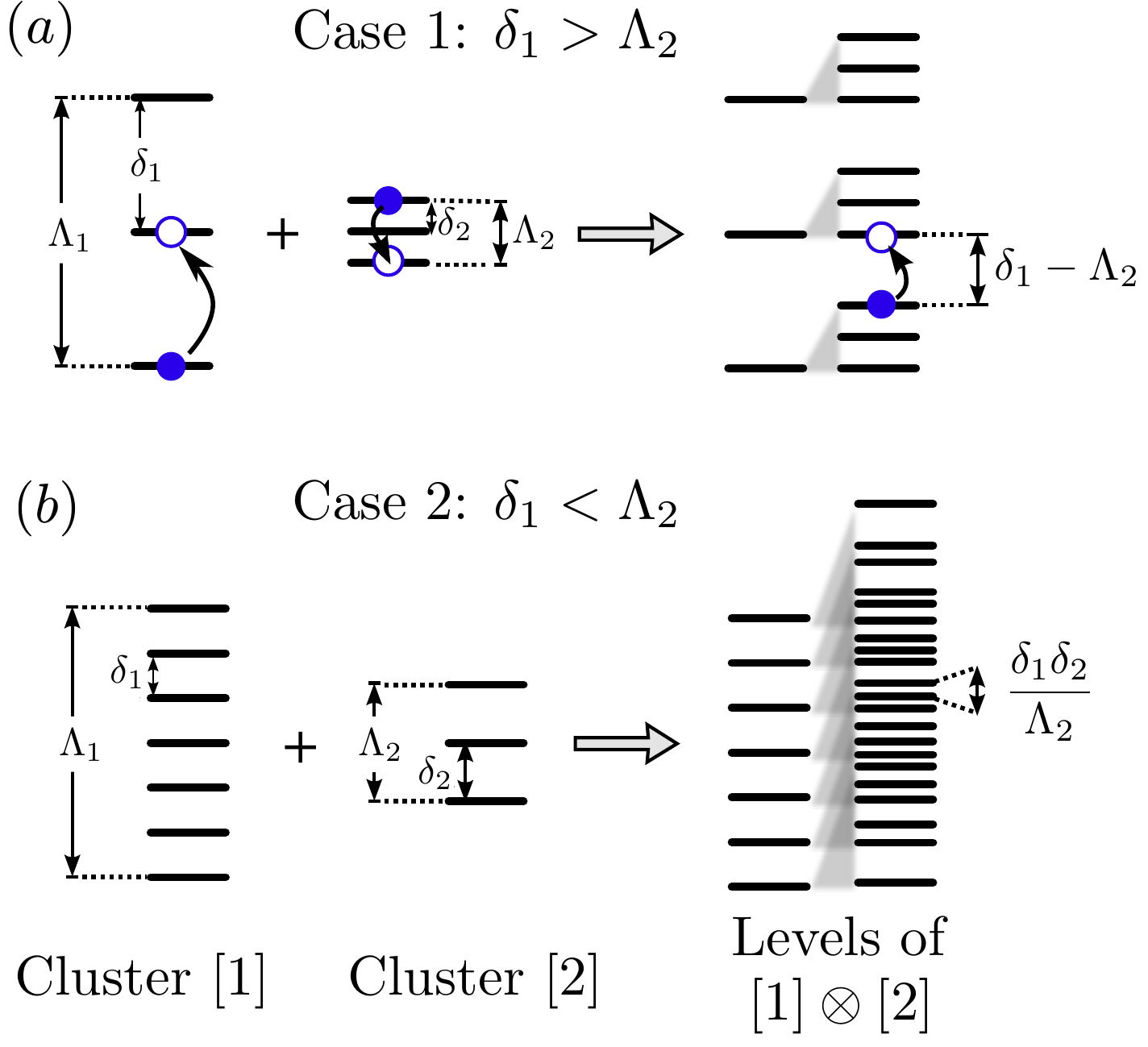}
\end{center}
\caption{{\bf Two cases for merging resonances.} The energy cost for changing the state of one cluster (blue circle to open circle connected by black arrow) can be partially compensated by changing the state of another cluster. The reduced energy penalty, $\delta E_{12}$, for such ``flip-flop" transitions is either $\delta_1-\Lambda_2$ or $\frac{\delta_1\delta_2}{\Lambda_2}$ depending on whether $\delta_{1}\gtrless\Lambda_2$, as shown in panels (a) and (b) respectively.}
\label{fig:ResMerging}
\end{figure}

\label{SecResMergingApp}

In the first stage of resonance formation, we check for pairwise resonances between two spin pairs $i$ and $j$ for which $\Gamma(x_{ij})> \delta E_{ij}$, where $\Gamma(x) \approx V e^{-x/x_0}$ is the typical effective coupling through the intervening strong disorder phase given by (\ref{eqJtypgen}). Here,  $x_{ij}$ is the distance separating the spin pairs $i,j$ and $\delta E_{ij}$ is their energy difference. As we iterate this procedure we generate larger clusters and check whether excitations can resonantly tunnel between any two clusters. Specifically, at any stage in the iterative process, we have a collection of $N$ resonant chains labeled by $j=1\dots N$, each comprised of $(m_j+1)$ bonds. Each resonant cluster has $\approx 2^{m_j}$ energy levels, spread over bandwidth $\Lambda_j$ equal to the sum of couplings $J$ for the resonant links inside the cluster, and hence have typical level spacing $\delta_a\approx \frac{\Lambda_j}{2^{m_j}}$. The energy levels within a given resonant chain are all strongly admixed and exhibit level repulsion.

If the residual coupling $\Gamma(x_{ij})$ between two clusters $i,j$ can resonantly drive transitions among the levels, then we merge these two resonances into a new, larger one. The precise rule for whether or not to merge two resonances depends on the relative size of the bandwidths and level spacings for each. Consider two resonant clusters, labeled $j=1,2$. Without loss of generality, assume that cluster $1$ has larger bandwidth $\Lambda_1>\Lambda_2$. There are two distinct cases to consider. If the level spacing of the first cluster exceeds the bandwidth of the second, $\delta_1> \Lambda_2$ (Fig.~\ref{fig:ResMerging}a), then the minimum energy difference associated with changing the states of both resonances is: $\delta E_{1,2} = \delta_1-\Lambda_2$. On the other hand, if $\delta_1<\Lambda_2$ (Fig.~\ref{fig:ResMerging}b), then the minimum energy cost for transitions is of order $\delta E_{1,2}\approx \frac{\delta_1\delta_2}{\Lambda_2}= \frac{\Lambda_1}{2^{m_1+m_2}}$.
 
If $\Gamma(x_{12})\ll \delta E_{1,2}$, then the residual coupling between the two resonant clusters merely weakly perturbs the energy levels of each. In this case, the combined energy levels of both clusters are well-described by the tensor product of the independent levels associated with each cluster indicating that energy excitations are well localized independently in each cluster. On the other hand, if $\Gamma(x_{12})\gg \delta E_{1,2}$, then the energy levels of both resonances are strongly admixed such that excitations are spread throughout both clusters, and energy can resonantly shuttle between them. In our numerical procedure, we take $\delta E$ as a sharp cutoff, merging resonances coupled by $\Gamma(x_{ij})> \delta E_{ij}$.

\subsection{Renormalization of inter-cluster couplings}
In this Appendix we explain the conceptual necessity of renormalizing the inter-cluster couplings in the effective resonance percolation model for the MBL transition. Perhaps surprisingly, we find that even though the renormalization steps are conceptually necessary to obtain a well-defined transition in asymptotically large systems, in practice, they may be omitted without changing the universal scaling properties obtained from finite-size scaling in moderately large systems.

If one omits the renormalization of inter-cluster couplings, and simply takes bare couplings: $\Gamma_{ij} = Ve^{-x_{ij}/x_0}$, where $x_{ij}$ is the distance between clusters $i$ and $j$, there is, in principle, a rare event driven breakdown of the strong-disorder phase which rounds the transition in the effective resonance percolation model~\cite{LaumannPC,HusePC}. Namely, consider the case where one finds a resonant cluster of $N_*$ spins, such that the bandwidth of this cluster $\Lambda_{N_*}$ is larger than the maximal energy mismatch between two spins ($\approx W$). Then, with simple geometrically dictated couplings that depend only on the inter-spin distance, this $N_0$ size cluster will deterministically absorb its neighbors, and continue to grow in this fashion until it encompasses all spins.  Such an ``avalanche" style breakdown occurs even for arbitrarily large disorder, as there is always a non-zero probability of finding such an $N_*$ spin ``seed" cluster.

Specifically, for very strong disorder, $N_*\approx \frac{W}{V}\gg 1$ spins are required to form a cluster whose bandwidth, $\Gamma_{N_*}\lesssim N_*V$, exceeds the energy mismatch $W$. The probability of finding such an $N_*(W)$ spin resonant cluster is $P_\text{avalanche}\approx e^{-W/V}$, indicating that an avalanche will occur for systems larger than $L_\text{avalanche}\approx e^{W/V}$. While this system size is extremely large for strong disorder $W\gg V$, it nevertheless indicates that the non-renormalized couplings lead to the unphysical prediction that the MBL phase will always be unstable to such an avalanche-style breakdown, and shows that the inter-cluster renormalization steps are necessary to obtain an asymptotically well-defined phase transition. This avalanche instability is eliminated by proper renormalization of inter-cluster couplings. 
Again working in the limit of strong disorder, we observe that, as the ``seed" cluster is formed, its couplings to neighboring spins will be typically renormalized downward by a factor of $\frac{V}{W}$ for each spin in the ``seed" cluster. Hence barring an extensive set of accidental resonances, which has vanishing probability for large systems, the avalanche will be suppressed for $\frac{V}{W}\ll 1$.

\begin{figure}[t!]
\begin{center}
\includegraphics[width = 1.0\columnwidth]{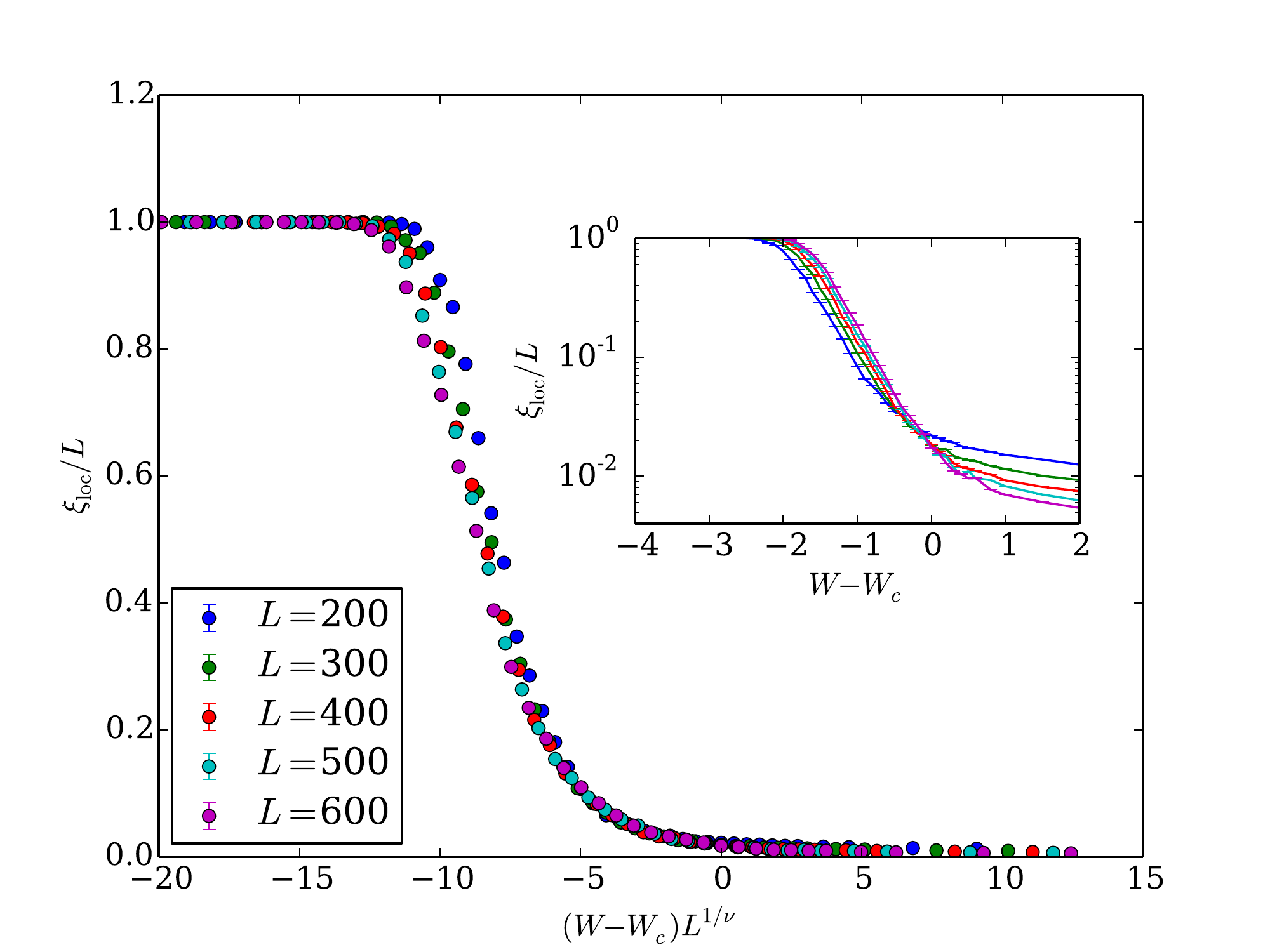}
\end{center}
\caption{{\bf Finite-size scaling for the disorder averaged localization length $\xi_{\text{loc}}$} (See Fig.~\ref{fig:XiScaling}) using a simplified model that ignores the renormalization of the couplings. Results are averaged over $\sim 10^4$ disorder realizations. 
}
\label{fig:XiScalingApp}
\end{figure}

On the other hand, simply omitting the renormalization steps and proceeding with the bare couplings, we obtain essentially identical finite-size scaling results compared to the case with renormalized couplings (Fig.~\ref{fig:XiScalingApp}). In particular we obtain finite size scaling curves for $\frac{\xi_\text{loc}}{L}$ that cross nicely at an apparent critical disorder strength $\frac{W_c}{V}\approx 17$ with $V = 0.3$. While the precise (non universal) value of $W_c$ differs in the absence of renormalization steps, the universal scaling structure such as the crossing value $\frac{\xi_\text{loc}}{L}\Big{|}_{W_c}\approx 2\cdot 10^{-2}$ and correlation length exponent $\nu\approx 3.5\pm 0.4$ agree within error bars with those results presented in the main text. 

Thus, the theoretically predicted avalanche phenomena appears to be completely absent even for relatively large ($L \approx 10^3$) system sizes. In fact, at the MBL transition point $\frac{W_c}{V}\approx 17$, from the arguments above, the avalanche instability would only be visible for system sizes of $L\gtrsim e^{17}$ spins, which is completely unaccessible by either numerical simulations or, for that matter, cold atoms experiments. Hence, while the renormalized couplings are {\it in principle} crucial for obtaining a well-defined transition, {\it in practice} one may omit the computationally expensive renormalization step and still observe universal critical finite-size scaling behavior in large systems. This situation is somewhat analogous to accessing critical properties of a conventional quantum critical point at very small, but finite temperature $T\approx e^{-W_C/V}$, which in cuts off the flow towards the universal scaling regime, but only above a very large length-scale, below which is it possible to accurately identify the universal scaling exponents.

\subsection{Construction of transport path}
Since we have seen that the potential avalanche instability is absent for computationally accessible system sizes at the critical point ($W=W_c$), we omit these computationally expensive steps in computing transport properties. However, in this case, some care is required to properly construct the transport path through the resonant cluster, and to avoid a related avalanche-style breakdown of the transport calculation for $W<W_c$. Namely, despite the absorption of all bonds into the resonant cluster for $W<W_c$, not all bonds contribute equally to transport, and some care is required to properly identify the transport path through the resonant cluster. An illustrative example is shown in Fig.~\ref{fig:MergingSequence}.

\begin{figure}[b!]
\begin{center}
\includegraphics[width = 1.0\columnwidth]{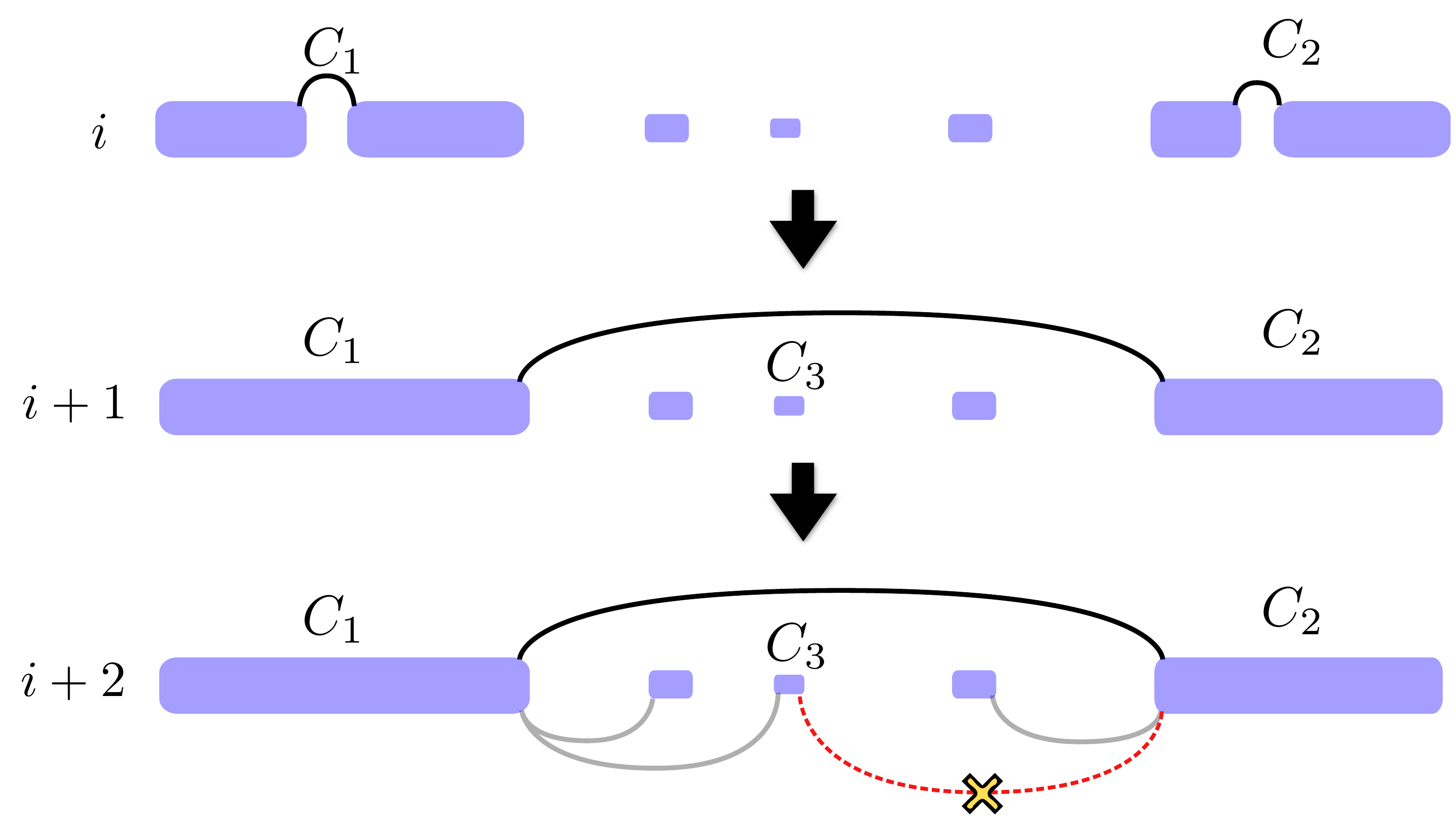}
\end{center}
\caption{{\bf Transport path through resonant cluster ($W<W_c$).} Schematic depiction of sequence of resonance-merging steps ($i,i+1,i+2$) that produces a gap in the transport path through the resonant cluster. See text for detailed explanation. Blue rectangles are resonantly linked clusters of spins. Black and gray lines indicate resonant links between previously formed clusters. Dashed crossed out red line indicates a spurious connection that should not be added to the transport path of the cluster.
}
\label{fig:MergingSequence}
\end{figure}

At some intermediate step, $i-1$, in the resonance-merging process small resonantly linked clusters of spins (blue rectangles) of various sizes have been identified. At step $i$, smaller clusters merge to form two large resonantly linked clusters, $C_{1},C_2$. In the next step, $i+1$, after accounting for the reduced level spacing of the newly formed clusters, one may find that $C_1$ and $C_2$ can resonantly exchange energy via the black link, and are merged into a single larger cluster $C_1\cup C_2$. It frequently happens that, at step $i+1$, the level spacing of $C_1$ (or $C_2$) alone is too large to facilitate the resonant exchange of energy between smaller clusters, e.g. $C_3$, between $C_1$ (or $C_2$), but that the combined level spacing of $C_1\cup C_2$ {\it is} small enough to resonate with $C_3$. Then, in the subsequent step, $i+2$, $C_1\cup C_2$ will absorb the intervening spins. However, it is crucial to note that, since $C_3$ cannot (in this example) resonantly exchange energy with $C_1$ or $C_2$ alone, but rather only by collectively exciting both $C_1$ and $C_2$. Hence, resonantly transferring energy between $C_3$ and $C_1\cup C_2$ takes {\it at least as long} as the time, $\tau(L_{12})$, to tunnel between $C_1\leftrightarrow C_2$ (long black link). A computationally inexpensive way to account for this feature, is to draw only a single connection between the closest pair of bonds on $C_3$ and $C_1\cup C_2$ (gray lines in Fig.~\ref{fig:MergingSequence}). Drawing additional connections, such as the crossed-out dashed red line in Fig.~\ref{fig:MergingSequence}, would spuriously allow energy to hop from $C_1\rightarrow C_3\rightarrow C_2$ in time $\tau(L_{13})+\tau(L_{32})\ll \tau(L_{12})$, whereas in this example we have seen that transferring an excitation from $C_3$ to $C_1\cup C_2$ must take at least $\tau(L_{12})$.

\end{appendix}

\end{document}